\documentclass[conference]{IEEEtran}

\usepackage{graphicx,times,amsmath}
\usepackage{mathtools}
\usepackage{subfigure}
\usepackage{color}
\usepackage{amsfonts}

\IEEEoverridecommandlockouts

\begin{document}

\title{\ \\ \LARGE\bf Investigating the Detection of Adverse Drug Events in a UK General Practice Electronic Health-Care Database \thanks{
$^{1}$ The Intelligent Modelling \& Analysis Research Group, School of Computer Science, The University of Nottingham, UK (email: {\tt jzr@cs.nott.ac.uk}).  $^{2}$ Division of Epidemiology \& Public Health, School of Community Health Sciences, The University of Nottingham, UK.}}

\author{Jenna Reps$^{1}$, Jan Feyereisl$^{1}$, Jonathan M. Garibaldi$^{1}$, Uwe Aickelin$^{1}$, Jack E. Gibson$^{2}$, Richard B. Hubbard$^{2}$}


\maketitle

\begin{abstract}
Data-mining techniques have frequently been developed for Spontaneous reporting databases.  These techniques aim to find adverse drug events accurately and efficiently.  Spontaneous reporting databases are prone to missing information, under reporting and incorrect entries.  This often results in a detection lag or prevents the detection of some adverse drug events.  These limitations do not occur in electronic health-care databases. In this paper, existing methods developed for spontaneous reporting databases are implemented on both a spontaneous reporting database and a general practice electronic health-care database and compared. The results suggests that the application of existing methods to the general practice database may help find signals that have gone undetected when using the spontaneous reporting system database.  In addition the general practice database provides far more supplementary information, that if incorporated in analysis could provide a wealth of information for identifying adverse events more accurately.
\end{abstract}


\section{Introduction}
The importance of data-mining techniques that could identify potential adverse drug events (ADEs) by analyzing information 
contained in large electronic medical databases was recognized decades ago.  The more available type of medical database, 
namely the spontaneous reporting system (SRS) database, often contains records for thousands of suspected adverse events.  
In SRS databases there is a connection in the database between any drug and suspected adverse event that is reported.  
The main problems that SRS databases face are incomplete or incorrect records, limited information and under-reporting \cite{Haramburu1993}. Existing data-mining techniques include the reporting odds ratio (ROR)\cite{Stricker1992}, the proportional reporting ratio (PRR)\cite{Evans2001}, the bayesian confidence propagation neural network (BCPNN)\cite{Bate2007} and the multi-item gamma poisson shrinker (MGPS) \cite{DuMouchel1999}.  These techniques have been developed specifically for use with the SRS databases.  They make use of the limited information by finding drug and adverse event associations that are disproportional relative to the rest of the drugs and events contained in the database.

A different type of medical database, referred to in this paper as the general practice (GP) database, contains the electronic records from a UK general practice. GP databases contain a greater depth of information for a patient, including details of all prescribed medications while the patient is registered. In general practice databases there are no direct drug 
and ADE connections.  These can only be predicted by looking for events that occur for a set time period after a drug is 
taken. However, events that occur after a drug is taken may be linked to the cause of taking the drug, may be age related 
or be seasonal effects.  This difference between the database structures may prevent the existing methods from being effective. Nevertheless, if the existing methods can be applied to the GP database successfully, this may enable the identification of real signals that were missed from the SRS databases due to incorrect records and under-reporting.

Recently there has been a focus on combining both the SRS and GP databases for signal detection \cite{Coloma2011}.  Little research has focused on developing or implementing an effective and efficient method for signal detection only on a GP database. In this paper the application of existing data-mining techniques, developed for the SRS database, to the GP database will be investigated. The ROR and PRR are implemented as these methods are more efficient when investigating a single drug. It has been shown in the past that if each drug-ADE has a frequency of four or more, all the methods have fairly similar results \cite{Puijenbroek2002}. As there is currently no golden standard for the SRS data-mining techniques, drugs with known side effects will be used for the investigation.


In the next section the standard data mining methods used in this paper are outlined along with an explanation of a data transformation approach that is used in order to make the GP database usable by the standard methods. A technique for comparing the obtained results, namely the receiver operating characteristic analysis, is then described. Subsequently a summary of the results obtained by the application of the existing methods on the GP and the SRS databases is presented. The paper concludes with a discussion of the obtained results and possible future work that warrants investigation due to our attained findings. 

\section{Methods}
In this work we are interested in comparing two of the standard techniques, namely the reporting odds ratio (ROR) and proportional reporting ratio (PRR) on two different datasets. Both methods use the values according to a $2$ $\times$ $2$ contingency table, shown in Table \ref{table:1}. 

\begin{table}
\centering
\begin{tabular}{ccc} \hline
 &Drug of interest&Other drugs\\ \hline
Event of interest& a& b \\
Other event & c & d\\
\hline\end{tabular}
\caption{Statistical significance in SRS databases are often calculated using
the frequencies $a,b,c,d \in \mathbb{Z}_{\geq 0}$.}
\label{table:1}
\end{table}

\subsection{Reporting Odds Ratio}
ROR is a ratio of two other ratios. It can be denoted as follows:

\begin{equation}
ROR = \frac{a/b}{c/d}
\label{eq:ror1}
\end{equation}

where values $a,b,c,d$ are calculated according to Table \ref{table:1}. The value ($a/b$) is the number of patients who had the event of interest and have taken the drug of interest (a) divided by the number of patients who had the event when taking any other drug (b). Thus this gives the ratio of the drug being taken, relative to all other drugs, for patients who had the event of interest. This is then compared to an analogous calculation for all other events, i.e. the number of patients who have taken the drug of interest but did not have the event of interest (c), divided by the number of people who had any other events, given that they took any other drug (d).

The standard error (SE) for the ROR method is calculated as follows:

\begin{equation}
SE(\ln ROR) = \sqrt{\frac{1}{a} +\frac{1}{b} + \frac{1}{c} + \frac{1}{d}}
\label{eq:ror2}
\end{equation}

where `ln' denotes the natural logarithm. This equation is used in the signal threshold calculation described below.

\subsection{Proportional Reporting Ratio}
The PRR on the other hand can be calculated as follows:

\begin{equation}
PRR = \frac{a/(a+c)}{b/(b+d)}
\label{eq:prr1}
\end{equation}

where $a/(a+c)$ can be thought of as the probability of an event of interest occurring, given the drug of interest was taken and an event occurred. $b/(b+d)$ can be thought of as the probability that the event of interest occurred, given any other drug was taken and an event occurred. Therefore the PRR approximates the ratio of the conditional probabilities for the event of interest, given the drug of interest and other drugs or given the other drug. 


The standard error (SE) for the PRR method is calculated as follows:

\begin{equation}
SE(\ln PRR) = \sqrt{\frac{1}{a}-\frac{1}{a+b}+\frac{1}{c}+\frac{1}{c+d}}
\label{eq:prr2}
\end{equation}

\subsection{Standard Error of ROR and PRR}

If the value of the ROR or PRR is larger than one, this suggests that there is a deviation from the background rates that are inferred by the frequency of events that occur for drugs other than the drug of interest. The signal thresholds for the ROR and PRR are $ROR-1.96 \times SE>1$ and $PRR-1.96 \times SE>1$ respectively \cite{Li2008}. These correspond to the lowest value of the $95\%$ confidence interval being greater than $1$.

\subsection{Database Pre-Processing}
\begin{figure}[t!]
\centering
\includegraphics[width=0.4\textwidth]{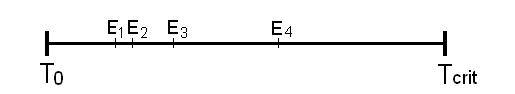}
\caption{A patient time-line starting from the time the drug is taken, $T_{0}$.  $T_{crit}$ is the time period chosen to find suspected drug-ADE relations and $E_{i} \; \forall i \in [1,n], n \in \mathbb{N}$ are events that occur between $T_{0}$ and $T_{crit}$.}
\label{figure:timeline}
\end{figure}

To calculate the ROR and PRR for the drugs and events in the GP database first requires identifying suspected drug-ADEs.  
This was done by finding events that occur within a set time period after a drug is taken for each patient. The time period used is $T_{crit}-T_{0}$ where $T_{0}$ is the time at which a drug is taken and $T_{crit}$ is a value defining the length of the time window. A graphical representation of this can be seen in Figure \ref{figure:timeline}.

Two values of $T_{crit}$ are investigated, $T_{crit}= 2$ months and $T_{crit}= \infty$.  The $T_{crit}= 2$ months was suggested by a medical expert to predict immediately occurring ADEs as people may not report events immediately but increasing the time period further will introduce more noise.  The $T_{crit}= \infty$ was investigated as this may help identify events that do not occur immediately. 

In the rest of this work, the dataset of potential adverse events for each drug using the association $T_{crit}= 2$ and $T_{crit}= \infty$ is defined as $GP_{2}$ and $GP_{\infty}$ respectively. The ROR and PRR and their respective standard errors are then calculated using Equations (\ref{eq:ror1}) to (\ref{eq:prr2}) for these two datasets.

\subsection{Receiver Operating Characteristic for the Comparison of Medical Databases}
The receiver operating characteristic (ROC) analysis was chosen to investigate the suitability of applying the existing methods to the GP database.  A ROC analysis plots the true positive rate (TPR) against the false positive rate (FPR), where these are defined in Equations (\ref{eq:roc1}) and (\ref{eq:roc2}). In our scenario the TPR is the amount of correctly identified adverse events divided by the number of adverse events. A FPR is one minus the number of correctly identified non-adverse events divided by the number of non-adverse events. The higher the value of the false positive rate, the more noisy the result. A method with such properties is naturally of little practical use.
\begin{equation}
\text{TPR} = \frac{A}{A+C} \\
\label{eq:roc1}
\end{equation}
\begin{equation}
\text{FPR} = 1-\frac{D}{B+D}
\label{eq:roc2}
\end{equation}
The existing methods have been successfully implemented on the SRS database in the past. By comparing the ROC plots of the methods on the SRS database with that of the GP database may give some indication of the suitability of applying the methods to the GP database. For each method the true positive rate and false positive rate  are calculated using the values describe in Table 2 over a range of signal threshold values.

\begin{table}
\centering
\begin{tabular}{cccc}\hline
 & Known  & Not known  & Total \\
&side effect&side effect& \\ \hline
Signaled & A & B & A+B \\
Not signaled & C & D & C+D \\
Total & A+C & B+D & N=A+B+C+D \\
\hline
\end{tabular}
\label{table:sspn}
\caption{Values used to calculate the true positive and false positive rates used in the ROC analysis.}
\end{table}

\subsubsection{Hierarchical Structures in Databases}
Some medical databases employ a structure in the way medical events are recorded. The GP database used in this work has a hierarchical structure for reported events. A typical event $E_j$ is represented by a code $x_{1}x_{2}x_{3}x_{4}x_{5}$ where $x_{i} \in \{0\} \cup \mathbb{N}_{\leq 9} \cup \{\mbox{a-z}\} \cup \{\mbox{A-Z}\} \cup \{\bullet\}$.  An example of an event code is ``\texttt{AB1a$\bullet$}''. The level of an event can be calculated as follows:
\begin{equation}
L(E) = 
	\begin{dcases}
		\operatorname*{arg\,min}_{i} \{{(i-1)|\; \;  x_{i}=\bullet}\}	&	\text{if } \exists \; i \; \mbox{s.t.} \; x_{i}=\bullet \\
		5	&	\text{otherwise}
	\end{dcases}
\end{equation}
The above given example event is reported at level 4. The higher the level, the greater the detail known about an event.  A level 4 event $x_{1}x_{2}x_{3}x_{4}\bullet$ is a more detailed version of a level 3 event $x_{1}x_{2}x_{3}\bullet\bullet$ .  It can also be seen that a patient who has an event $x_{1}x_{2}x_{3}x_{4}\bullet$ also must have the event $x_{1}x_{2}x_{3}\bullet\bullet$ .

The ROC analysis requires the knowledge of known adverse events.  The known adverse events used are those listed as possible side effects in the BNF \cite{bnf1}.  For each of the known adverse events, the set of event codes relating to event $E_j$ is found and denoted $K_{j}$. For example if an event of interest is ``nausea'', all event codes related to this event are contained in the set $K_{1}$. These event codes however also range over the hierarchal levels. It is thus common in the GP database for each of the known adverse events, $K_{j}, \; j \in [1,n]$, where $n$ denotes the number of adverse events, to have different cardinalities. 

The function $I$ defined below finds the number of known adverse events that have at least one of the associated event codes signaled by the data-mining technique. The purpose of the $I$ function is to find a fair way of determining the TPR of the methods when each $K_j$ have different cardinalities.
\begin{equation}
I(K_{j}) = \left\{
  \begin{array}{l l}
    1 & \quad \text{if } K_{j} \cap S \neq \emptyset \\
    0 & \quad \text{if } K_{j} \cap S = \emptyset \\ 
  \end{array} \right.    
\end{equation}
Letting $K= \bigcup_{j}K_{j}$ denote the set of all known adverse event codes, $S$ denotes the set of all signaled event codes, 
$\Omega$ the set of all event codes and  $|\;|$ the cardinality of a set. 

The values $A,B,C$ and $D$ in Table 2 are found by:
\begin{align}
A & = \sum_{j=1}^{n}{I(K_{j})} \\
B & = |S \cap K^{c}| \\
C & = n - \sum_{j=1}^{n}{I(K_{j})} 
\end{align}

\begin{equation}
D = |S^{c} \cap K^{c}|
\end{equation}

where $K^c$ is the complement of $K$. This is equal to $\Omega \setminus K$, which can be stated as the sample space less $K$. $S^c$ is analogous to $K^c$.

\section{Experimental Setup}
In the performed experiments two different databases, GP Database and SRS Database, are investigated using the two different methods, namely ROR and PRR.

\subsection{General Practice Database}
The two databases investigated in this paper are the GP database containing complete records of patients while registered at a UK general practice.  These records include medical events, drugs prescribed, family histories, general demographics and patient information.  The patient information contains useful data such as date of birth, gender, date of registration and if present, date of death.  Every time a patient visits the surgery a record will be added to the database regarding events relating to that visit.  Entering the data is compulsory. The GP database studied contains medical records for a total of 69616 patients. These records span over a 107 year time-period from 1902 until 2010 and contain 1,858,229 medical events and 678,159 drug prescriptions.

\subsection{Spontaneous Reporting System Database}
The SRS database studied in this paper is the US Food and Drug Administration (FDA) Adverse Event Reporting System (AERS).  The database contains information on possible drug-ADE interactions. Included in the database is information regarding the drug taken, the possible adverse drug event, the general outcome of the drug and the patient information, including demographics. The data is recorded by professionals or the general public when a suspected drug-ADE has occurred, however not all information is always included.  Often records are incomplete and mistakes can be made.  This type of database is also prone to under reporting.  The time period for the records investigated range from first quarter of 2004 to the second quarter of 2009.  The total number of drugs sequences reported is 6,354,539 and the total number of unique events names recorded is 14,064. For the general patient information 51\% of the records contain missing information either regarding the patients age, gender or event date.  Gender is only recorded 94\% of the time and age 64\% of the time.   

\subsection{Tested Hypotheses}
The following null hypotheses, along with their alternatives, are investigated in this paper:

{\bf ROR/PRR ROC (Segment) Comparison:}\\
\hangindent=0.3cm \emph{H$_{0}$}: There is no statistically significant difference between the areas under the receiver operating characteristic curve (for the FPR less than \emph{n}\%) calculated by applying the ROR/PRR on the GP and SRS databases.\\
\emph{H$_{1}$}: There is a statistically significant difference between the areas under the receiver operating characteristic curve (for the FPR less than \emph{n}\%) calculated by applying the ROR/PRR on the GP and SRS databases.\\

The ROC segment comparison is investigated ($n<100$) as the method performance is only relevant for low FPR's. A large FPR means that the actual adverse events found would be swamped by an excess amount of non-adverse events that are essentially noise.  As the maximum number of events in the three datasets is approximately $14,000$, the value $n=0.02$ is investigated. This means the maximum number of false positive signals is less than 300 and therefore less that ten times the number of actual adverse events for one of the investigated drugs. 

The hypotheses are tested by first mapping the GP data into a structure suitable for the application of the ROR and PRR methods. Different criteria for the mapping are investigated. The ROR and PRR are performed on both databases for two chosen drugs of interest.  The drugs of interest are Ethinylestradiol and Amoxicillin.

Once the ROR and PRR are calculated a ROC plot is produced by investigating the true positive and true negative rates form the signals produced using the thresholds: $ROR \geq n$, $n \in [0,\infty)$ and $PRR \geq n$, $n \in [0,\infty)$.  This shows the trade off between the number of true positives against the number of false positives as the threshold becomes less stringent.

The area under the curve (AUC) is calculated numerically for all ROC plots. The AUC's for the ROR and PRR applied to the different databases are then tested for significant differences by the methods described in \cite{Hanley1983}. The statistical analysis for the AUC of the ROC segment is calculated in a similar manner.




\subsection{Investigated Drugs}
\subsubsection{Ethinylestradiol}
Ethinylestradiol is a commonly used drug found in the combined oral contraceptive pill.  Ethinylestradiol has been around for more than half a century and the possible side effects have been well studied \cite{bnf1}. Recently there has been interest in a possible connection between Ethinylestradiol and idiopathic venous thromboembolic events \cite{Girolami2002}.

\subsubsection{Amoxicillin}
Amoxicillin is a frequently used penicillin based antibiotic that had the greatest number of prescriptions within the GP database. It has been in use for almost four decades \cite{ Bayne1974}. As these drugs have been around for many years and are frequently used, known side effects have been observed and reported, including: nausea, vomiting, diarrhoea, rashes and less commonly, antibiotic-associated colitis \cite{bnf2}.  

\section{Results}
A total of 14,064, 10,755 and 9,158 possible adverse events were found in the SRS database, GP with $T_{crit}=\infty$ and GP with $T_{crit}=2$ respectively.

\subsection{Ethinylestradiol}

%


\begin{figure*}
\vspace{-8mm}
\centering
\subfigure[ROC analysis for ROR]{
	\includegraphics[width=0.4\textwidth, height=0.4\textwidth]{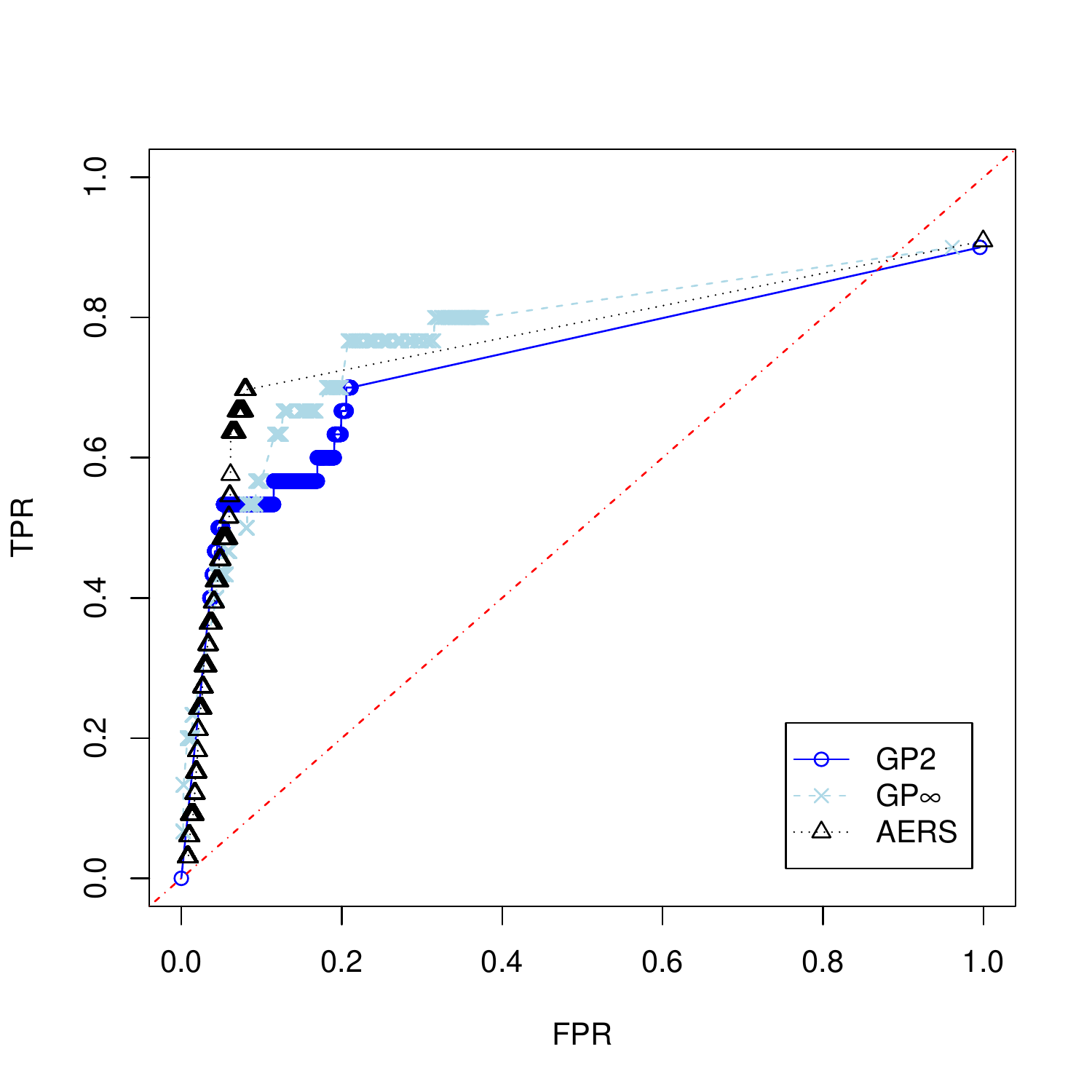}
	\label{fig:ethinyROR}
}
\subfigure[ROC analysis of PRR]{
	\includegraphics[width=0.4\textwidth, height=0.4\textwidth]{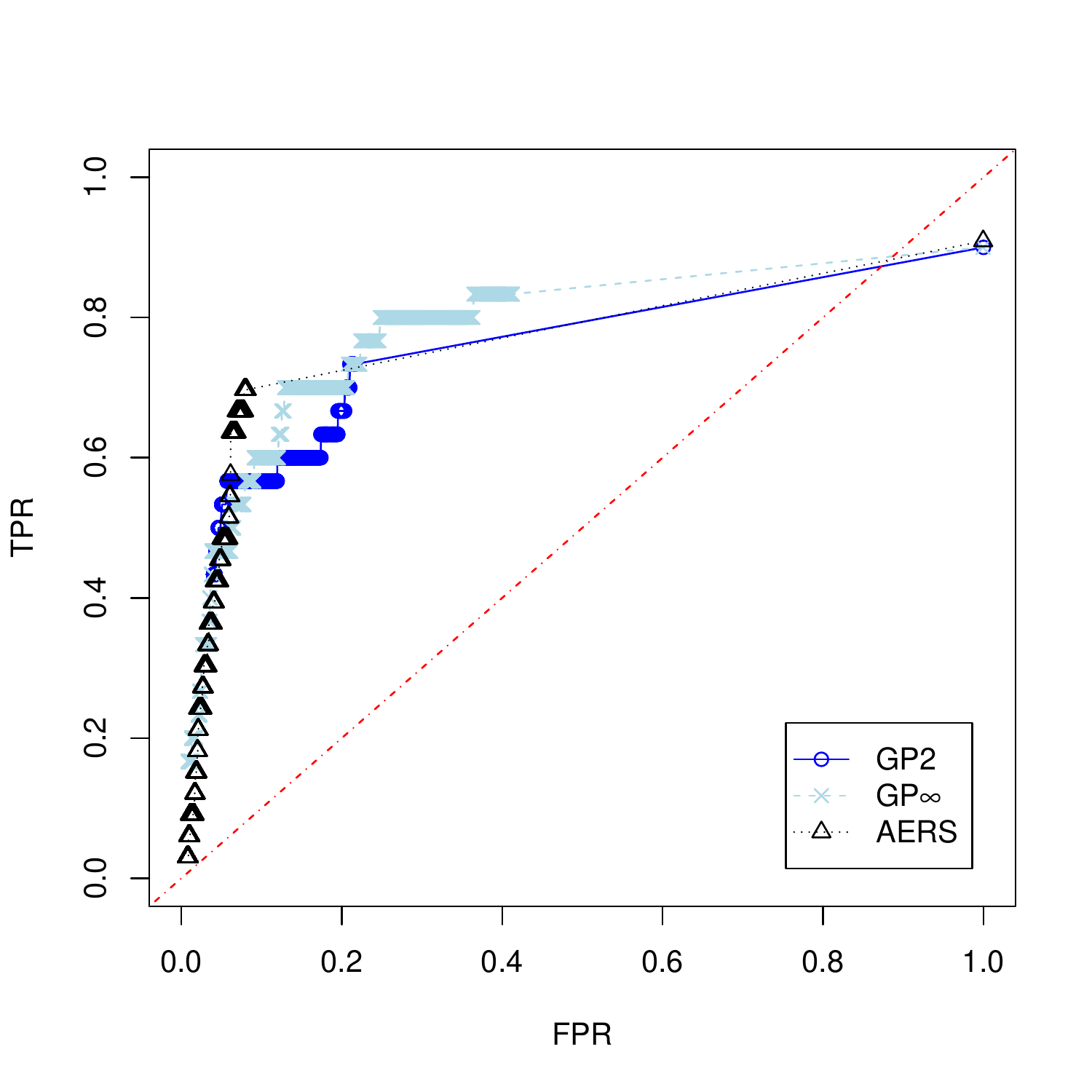}
	\label{fig:ethinyPRR}
}
\caption{ROC analysis of Ethinylestradiol. Results for both the ROR and the PRR methods on the three tested datasets are shown.}
\end{figure*}



The ROC plots, shown in Figures \ref{fig:ethinyROR} and \ref{fig:ethinyPRR}, illustrate that the maximum TPR for both the ROR and PRR are similar when applied to the three datasets.  It is also apparent that the TPR increases rapidly for all three datasets between FPR values 0 and 0.1. Between FPR values 0.1 and 0.2 the methods applied to the SRS database give the greatest TPR. For values of FPR greater than 0.3 the database returning the highest values of TPR is $GP_{\infty}$.

The AUC over the whole range of FPR for the application of the ROR method with their corresponding $95\%$ confidence interval are: 0.73 (0.63-0.83), 0.77 (0.68-0.87) and 0.77 (0.67-0.86) for $GP_{2}, GP_{\infty}$ and SRS respectively.  Therefore overall the ROC performed the worse on the $GP_{2}$ dataset.  The AUC values for the $GP_{2}$ and $GP_{\infty}$  were not significantly different to that of the SRS, with corresponding p-values of 0.31 and 0.46. Leading to the rejection of the alternative hypothesis.

The PRR gave similar results of 0.74 (0.70-0.79), 0.79 (0.69-0.88) and 0.77 (0.67-0.86) for the $GP_{2}, GP_{\infty}$ and SRS respectively. Again, the AUC values for the general practice datasets where not significantly different to the AUC value of the SRS.


The AUC values and the $95\%$ confidence intervals for the ROR over the FPR range of 0 to $0.02$ were 0.0034 (0.0020-0.0048) and 0.0012 (0.0006- 0.0017) for the $GP_{\infty}$ and SRS respectively. The test for statistical similarity returned a p-value of $0.002$.  Therefore we reject the null hypothesis that the AUC values are the same between the GP dataset and the SRS dataset. The corresponding values for the PRR method were 0.0021 (0.0011- 0.0030) and 0.0002 (3.2 $\times 10^{-6}$ - 0.0003) for the $GP_{\infty}$ and SRS respectively.  The p-value for the statistical test for similarity between the two AUC values is $0.00058$.  Again, this leads to the rejection of the null hypothesis that the AUC values obtained by applying the methods to the GP and SRS database are the same for small FPRs.    

When applying the general signal threshold criteria, $ROR-1.96 \times SE>1$ or $PRR-1.96 \times SE>1$, the ROR and PRR applied to the GP database with $T_{crit}=\infty$ found the highest number of known adverse drug events, 17 and 19 respectively.

\subsection{Amoxicillin}

\begin{figure}
\vspace{-8mm}
	\centering
		\includegraphics[width=0.4\textwidth, height=0.4\textwidth]{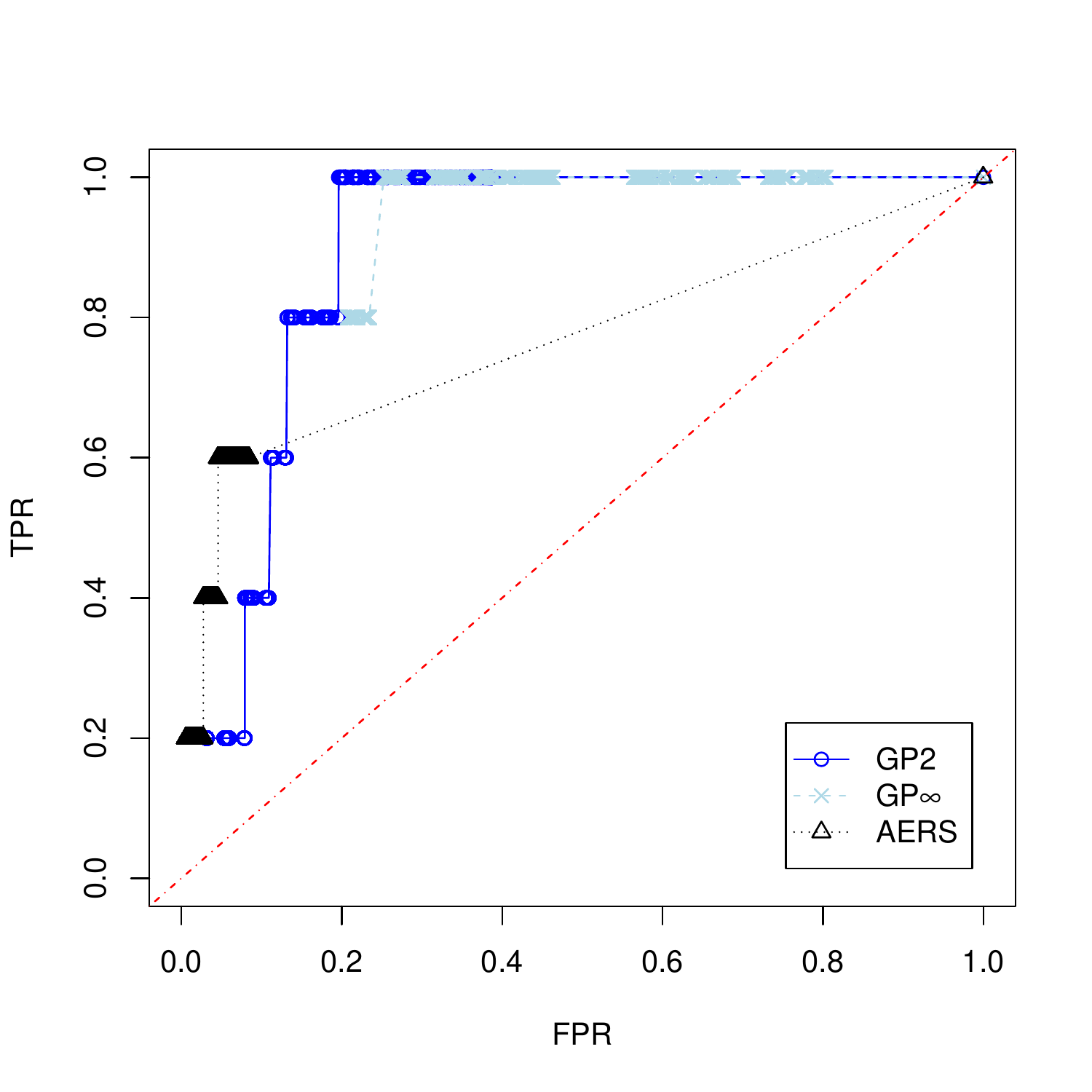}
		\label{fig:amoxROR}
		\vspace{-2mm}
	\caption{The Amoxicillin ROC analysis for the ROR. It is apparent that the standard methods applied to the SRS database produce better results when the FPR is low, however with this database it is impossible for the standard methods to achieve high levels of TPR. The PRR produced visually identical results.}
\end{figure}


The Amoxicillin ROC analysis for both the ROR and PRR is shown in Figures \ref{fig:amoxROR} and \ref{fig:amoxPRR}. The PRR and ROR gave similar ROC plots for Amoxicillin. For this drug, in all cases, the null hypothesis could not be rejected. This implies that statistically there exists no significant difference between the results. From the figures it can be seen that for low values of FPR the standard methods applied to the SRS dataset give better results than for the GP database, however, using the SRS dataset the methods cannot detect all adverse events correctly. On the contrary, the GP dataset allows for the correct detection of true adverse events whilst the number FPR value increases.

Overall the $GP_{2}$ dataset had a greatest AUC. Applying the existing methods to the SRS dataset gave the worse results in terms of known adverse drug events found as can be seen in Table \ref{tab:roc_amox}. Using the standard signal criteria, $ROR-1.96 \times SE>1$ or $PRR-1.96 \times SE>1$, the PRR applied to the $GP_{\infty}$ dataset however detected all five known adverse events.

\begin{table*}[t!]
\centering
\begin{tabular*}{0.8\textwidth}{ccccccc}\hline
Adverse Event & \multicolumn{2}{c}{ROR (GP)} & ROR (SRS) & \multicolumn{2}{c}{PRR (GP)} & PRR (SRS)\\
              & $T_{crit}=2$ & $T_{crit}=\infty$ &         &$T_{crit}=2$ & $T_{crit}=\infty$&          \\
\hline
Nausea &$\times$&$\times$ & $\times$ & $\times$ & \checkmark & $\times$ \\
Vomiting &\checkmark &\checkmark & \checkmark & \checkmark &\checkmark  & \checkmark \\ 
Diarrhoea &\checkmark &\checkmark & $\times$ & \checkmark &\checkmark  & $\times$\\ 
Rash  &\checkmark &\checkmark & \checkmark & \checkmark &\checkmark  & \checkmark \\ 
Colitis  & $\times$ & \checkmark & \checkmark & $\times$ & \checkmark & \checkmark \\
\hline
\end{tabular*}
\caption{Table indicating if the method was able to detect known adverse events of Amoxicillin with signal criteria $ROR-1.96 \times SE>1$ or $PRR-1.96 \times SE>1$.}
\label{tab:roc_amox}
\end{table*}


\section{Discussion}
The better results obtained from the standard methods applied to the GP datasets over the SRS database in the segment ROC analysis for Ethinylestradiol suggests that the GP database contains at least as good information as the SRS database in this case. The result for Amoxicillin on the other hand reveals an interesting phenomenon of the GP database. The SRS database does not contain enough information in order to allow the standard methods to successfully detect all potential adverse events. The GP datasets on the other hand allow for the detection of all adverse events, however only under the condition of having high FPR. This suggests that the broader scope of the GP database offers potential benefits, however in order to be able to obtain these, methods which can reduce FPR need to be developed.

The result in the Amoxicilin case may be due to the way potential adverse events are found via the GP database. As mentioned in the introduction, some events that occur frequently after the drug is taken may be related to the event that caused the drug or be a seasonal effect. This could cause the frequency of the specific drug and event to be higher relative to the rest of the drugs and events, even though they are not directly linked. Because of this, the ROR may produce many false positives. Ethinylestradiol is frequently taken as a monthly repeat prescription. This means the time window that events must occur within, to be considered potentially adverse, can last for years. This longer time window may help to average out the effects of events indirectly linked to the drug and therefore reduces the amount of false positives. 

The obtained results show that applying the ROR and PRR on the GP database may help identify new signals. The known adverse events detected for each drug differ between all three datasets.  Many of the known adverse events were unable to be identified using the standard signal criteria for the SRS.  However, these were often found when using the GP database.  It is worth mentioning that although applying the existing methods to the GP database for Ethinylestradiol resulted in more known adverse event signals, as mentioned above there were also more false positive signals.  As the ROC plots show all three datasets have similar sensitivity and specificity associations. This suggests a different signal criteria needs to be implemented when applying the existing methods to the GP database.

The current method of applying the existing data-mining techniques to the GP database did not consider the potential benefit of having a multiple level hierarchy for the event codes.  Rather than calculating the ROR for all the event codes, it may be beneficial to apply the ROR separately for each of the levels. It may be common for doctors to record events as less specific than actually known.  This may prevent signals for specific adverse events being detected.  For example, the fifth level adverse event `reduced menstrual loss' may be frequently recorded as the fourth level `change in menstrual loss'. The method implemented in this paper would not have detected `change in menstrual loss' as a known signal. The reason being, higher level event codes of known adverse events were not considered to be known adverse event codes. However, calculating the ROR for level four event codes may give a signal for `change in menstrual loss' which could then be considered a possible signal for related level five event codes such as `reduced menstrual loss'.  These level five event codes could then be further investigated with a case control analysis. 

\section{Conclusions}
In this work we have investigated the applicability of two standard data-mining techniques for the detection of potential adverse drug events in two different medical databases. The spontaneous reporting system database provides a library of suspected adverse events, however with limited information and issues, such as under-reporting. Methods developed for this type of database are limited by the quality and scope of the data that the database provides. On the other hand the general practice database provides a more comprehensive medical record of patients that includes complete patient history and thus provides greater possibilities for finding true potential adverse events. Initially a method for transforming the general practice database into a form processable by the standard techniques has been proposed. Statistical tests have been subsequently used to investigate whether results obtained on the two different datasets are the same. The null hypothesis in the case of comparing the entire receiver operating characteristic curves was not rejected, however looking at the vital part of the curve, with false positive rates less than $2\%$, the null hypothesis was rejected and it has been shown that the standard methods can perform better on the general practice database. This suggests that new techniques need to be developed in order to fully utilize the information contained within the general practice database, resulting in potentially more accurate adverse drug event reporting. Beside the utilization of a wider variety of existing information in the general practice database for analysis, the hierarchical structure of the event coding system should be considered in order to help with issues such as under-reporting and missing information.

\bibliographystyle{abbrv}
\bibliography{mybib}  

\end{document}